\documentclass[12pt]{article}

\usepackage{bm}
\usepackage{color}
\usepackage{amsmath}
\usepackage{amssymb}
\usepackage{amsfonts}
\usepackage{times}
\usepackage{graphicx}

\usepackage[numbers,sort&compress]{natbib}

\newenvironment{sciabstract}{%
\begin{quote} \bf}
{\end{quote}}

\title{Particle Production via Dirac Dipole Moments in the Magnetized and Non-magnetized Exponentially Expanding Universe}

\author
{Semra Gurtas Dogan$^{1}$, Ganim Gecim$^{2}$,  Yusuf Sucu$^{1,\ast}$\\
\\
\normalsize{$^{1}$Department of Physics, Faculty of Science, Akdeniz University, 07058 Antalya,
Turkey}\\
\normalsize{$^{2}$Department of Astronomy and Astrophysics, Faculty of Science, Atat\"{u}rk Univ.,25240 Erzurum,
Turkey}\\
\normalsize{$^\ast$Correspondence should be addressed; E-mail: ysucu@akdeniz.edu.tr}
}

\date{}

\begin{document}

% Double-space the manuscript.

\baselineskip24pt

% Make the title.

\maketitle

% Place your abstract within the special {sciabstract} environment.

\begin{sciabstract}
In the present paper, we solve the Dirac equation in the
2+1 dimensional exponentially expanding magnetized by uniform
magnetic field and non-magnetized universes, separately. Asymptotic
behaviors of the solutions are determined. Using these results we
discuss the current of a Dirac particle to discuss the polarization
densities and the magnetization density in the context of Gordon
decomposition method. In this work we also calculate the total
polarization and magnetization, to investigate that the magnetic
field how can effect on the particle production. Furthermore, the
electric and the magnetic dipole moments calculated, and based on
these we have discussed the effects of the dipole moments on the
charge distribution of the universe and its conductivity for both
the early and the future time epoch in the presence/absence a
constant magnetic field and exponentially expanding spacetime.
\end{sciabstract}

\section*{Introduction}
One of the most interesting and important results of the
formulation of the relativistic quantum mechanics in curved spacetime is the
particle creation event in the expanding universe which was firstly
discussed by Parker for the scalar particles and Dirac particles \cite{1,2,3}. So, he computed the number density of the created particles by means of
the Bogolibov transformation by using the out vacuum states constructed from the solutions of the relativistic particles wave
equations. After these important works of Parker, the solutions of the
relativistic particle wave equations have extensively been studied in
various 3+1 dimensional spacetime backgrounds \cite{4,5,6,7,7a,8,9,9a,9b,9c,9d,H1,H2,H3,H4,H5}. Using the WKB approach, the number
density and renormalized energy-momentum tensor of the created spin-1/2 particle in the spatially flat (3+1)-dimensional Friedmann-Robertson-Walker(FRW) spacetime have been calculated \cite{Ghosh}. The effect of the scalar particle creation on the collapse of a spherically symmetric massive star was investigated and it was demonstrated that the collapsing process was not independent from the particle creation rate \cite{Bhat}. Moroever, in \cite{Jawad} thermodynamics laws and equilibrium conditions are discussed in the presence of particle creation in the context of the (3+1)-dimensional Chern-Simons gravity theory. Recently, creation of the massless fermion in the Bianchi type-I spacetime investigated and shown that the massless particles can be created during the early anisotropic expansion epoch \cite{Bhoonah}.

The Gordon decomposition of the Dirac currents is another useful tool for
discussing the particle creation phenomena \cite{4,8,9,23}. In the
decomposition method, the Dirac currents constructed from the solutions of
the Dirac equations are separated into three parts, the convective,
polarization with three components, and magnetization with three components,
in the 3+1 dimensional spacetime. This method includes some complexities
stemming from the 3+1 dimensional spacetime. Using this method in a 2+1
dimensional curved spacetime, the densities of the particle currents are
separated into three parts, as in the 3+1 dimensional spacetime, but
polarization density has two components and the magnetization density has
only one component \cite{23}, and, moreover, as the Dirac spinor can be
defined by only two components, the computations in the 2+1 dimensional
spacetime become more simple than that of 3+1 dimensional spacetime. Because
of the simplicity stemming from the dimensions, the dipole moments that are
computed from the polarization and magnetization densities of the Dirac
electron under influence in a constant magnetic field are easily computed
and their result are, furthermore, compatible with the current experimental
results \cite{23a}. With these motivations, in this study, we solve the
Dirac equation in the 2+1 dimensional exponentially expanding magnetized by
uniform magnetic field and non-magnetized universes, separately, and discuss
the particle creation event by means of the Dirac currents written in terms
of these solutions. As a result, we observe that the polarization and
magnetization parts of the currents are affected differently whether the
exponentially expanding universe is magnetized or not and find expressions
for the electric and magnetic dipole moments by integrating the polarization
and magnetization densities on hypersurface.

The outline of the work is as follows; in Section 2, we, at first, discuss
the Dirac equation solutions in the 2+1 dimensional exponentially expanding
universe. In Section 3, the Dirac equation is solved in the 2+1 dimensional
exponentially expanding universe with a constant magnetic field. In Section
4, we derive the components of Dirac currents for the solutions obtained in
the Sections 2 and 3, and also compute the polarization density, the
magnetization density, total polarization (electric dipole moment) and the
total magnetization (magnetic dipole moment). Finally, last Section,
conclusion, includes a discussion about the results of this work.

\section{Dirac particle in the 2+1 dimensional exponentially expanding universe}

The behavior of the electron in 2+1 dimensional curved space is represented
by the covariant form of the Dirac equation \cite{23}, which is important application in curved spacetime \cite{a1,a2,a3,a4,a5,a6,a7,a8}
\begin{equation}
\left\{ i\overline{\sigma }^{\nu }(x)\left[ \partial _{\nu }-\Gamma _{\nu
}+ieA_{\nu }\right] \right\} \Psi (x)=m\Psi (x),  \label{Equation1}
\end{equation}%
where $\Psi(x)$ =$\binom{\Psi_{1}}{\Psi_{2}}$ is the Dirac spinorial wave
function with two components that are positive and negative energy
eigenstates, $m$ is the mass of Dirac particle, $e$ is the charge of the
Dirac particle and $A_{\nu}$ are 3-vectors of electromagnetic potential.
Using triads, $e_{(i)}^{\nu}(x)$, Dirac matrices that dependent on
spacetime, $\overline{\sigma}^{\nu }(x)$, are written in terms of the
constant Dirac matrices, $\overline{\sigma}^{i}$;
\begin{equation}
\overline{\sigma }^{\nu }(x)=e_{(i)}^{\nu}(x)\overline{\sigma}^{i}.
\label{Equation3}
\end{equation}
So, we choose the constant Dirac matrices, $\overline{\sigma }^{i},$ in the
flat spacetime as follows:
\begin{equation}
\overline{\sigma }^{i}=(\overline{\sigma }^{0},\overline{\sigma }^{1},%
\overline{\sigma }^{2})  \label{Equation4}
\end{equation}%
with
\begin{equation}
\overline{\sigma }^{0}=\sigma ^{3},\ \ \overline{\sigma }^{1}=i\sigma ^{1},\
\ \overline{\sigma }^{2}=i\sigma ^{2},  \label{Equation5}
\end{equation}%
where $\sigma ^{1}$, $\sigma ^{2}$ and $\sigma ^{3}$ are Pauli matrices. The
spin connection, $\Gamma _{\nu }(x),$ for the diagonal metrics \ is defined
as;
\begin{equation}
\Gamma _{\nu }(x)=-\frac{1}{4}g_{\tau \alpha }\Gamma _{\nu \beta }^{\alpha }%
\left[ \overline{\sigma }^{\tau }(x),\overline{\sigma }^{\beta }(x)\right] ,
\label{Equation7}
\end{equation}%
where $\Gamma _{\nu \beta}^{\alpha}$ is the Christoffel symbol and it gives as follows \cite{Weinberg}:
\begin{equation}
\Gamma_{\nu\beta}^{\alpha}=\frac{1}{2}g^{\lambda\alpha}\left[\partial g_{\beta\lambda,\nu}+\partial g_{\nu\lambda,\beta}-\partial g_{\beta\nu,\lambda}\right].
\label{Equation7a}
\end{equation}
Also, the metric tensor $g_{\beta \nu }(x)$ is written in terms of triads as follows;
\begin{equation}
g_{\beta \nu }(x)=e_{\beta }^{(i)}(x)e_{\nu }^{(j)}(x)\eta _{(i)(j)},
\label{Equation6}
\end{equation}%
where $\beta $ and $\nu $ are curved spacetime indices run from $0$ to $2$, $%
i$ and $j$ are flat spacetime indices run $0$ to $2$ and $\eta _{(i)(j)}$ is
the signature with (1,-1,-1).

The (2+1) dimensional de Sitter space-time metric can be written as \cite{24};
\begin{equation}
ds^{2}=dt^{2}-e^{2Ht}\left[ dr^{2}+r^{2}d\phi ^{2}\right] \
\label{Equation2}
\end{equation}%
where $H$ is Hubble parameter. From the Eqs.(\ref{Equation4})-(\ref{Equation2}), the spin connections for
the metric read
\begin{equation}
\Gamma _{0}=0\ ,\ \ \Gamma _{1}=-\frac{H}{2}e^{Ht}\overline{\sigma }^{0}%
\overline{\sigma }^{1}\text{ and }\Gamma _{2}=-\frac{1}{2}[rHe^{Ht}\overline{%
\sigma }^{0}\overline{\sigma }^{2}+\overline{\sigma }^{1}\overline{\sigma }%
^{2}].  \label{Equation10}
\end{equation}%
Using the Eqs.(\ref{Equation3}), (\ref{Equation2}) and
(\ref{Equation10}), then, the Dirac equation in the 2+1 dimensional
exponentially expanding universe becomes,
\begin{equation}
\big\{\overline{\sigma }^{0}\left( \partial _{t}+H\right) +im+e^{-Ht}[%
\overline{\sigma }^{1}(\partial _{r}+\frac{1}{2r})+\frac{\overline{\sigma }%
^{2}}{r}\partial _{\phi }]\big\}\Psi (x)=0.  \label{Equation12}
\end{equation}%
Letting the Eqs.(\ref{Equation5}) and (\ref{Equation12}) and the Dirac spinor, $\Psi (x)$ =$%
\binom{\Psi _{1}}{\Psi _{2}},$\ we write the Dirac equation in
explicit form as follows:
\begin{eqnarray}
\lbrack \partial _{t}+im+H]\psi _{1}+ie^{-Ht}[\partial _{r}+\frac{1}{2r}-%
\frac{i}{r}\partial _{\phi }]\psi _{2} &=&0  \notag \\
\lbrack \partial _{t}-im+H]\psi _{2}-ie^{-Ht}[\partial _{r}+\frac{1}{2r}+%
\frac{i}{r}\partial _{\phi }]\psi _{1} &=&0\ .  \label{Equation13}
\end{eqnarray}%
To find the solutions of Eqs.(\ref{Equation13}), thanks to the separation of
variables method, the wave function components can be defined as
\begin{equation}
\binom{\Psi _{1}}{\Psi _{2}}=e^{ik\phi }\binom{T_{1}(t)R_{1}(r)}{%
T_{2}(t)R_{2}(r)}.  \label{Equation14}
\end{equation}%
By these definitions, the Dirac equation are separated into the
following two differential equation systems:
\begin{eqnarray}
\left[ \frac{d}{dr}+\frac{1}{2r}+\frac{k}{r}\right] R_{2}(r) &=&-\lambda
R_{1}(r)  \notag \\
\left[ \frac{d}{dr}+\frac{1}{2r}-\frac{k}{r}\right] R_{1}(r) &=&\lambda
R_{2}(r)  \label{Equation15}
\end{eqnarray}%
and
\begin{eqnarray}
\left[ \frac{d}{dt}+im+H\right] T_{1}(t) &=&i\lambda e^{-Ht}T_{2}(t)  \notag
\\
\left[ \frac{d}{dt}-im+H\right] T_{2}(t) &=&i\lambda e^{-Ht}T_{1}(t),
\label{Equation16}
\end{eqnarray}%
where $\lambda $ is a separation constant, and we find the solutions of Eqs.(%
\ref{Equation15}) in terms of the Bessel and confluent hypergeometric
functions as follows:
\begin{eqnarray}
R_{1}(r) &=&AJ_{k-\frac{1}{2}}\left( \lambda r\right) =A\frac{e^{-i\lambda
r}\left( \frac{\lambda r}{2}\right) ^{k-\frac{1}{2}}}{\Gamma \left( k+\frac{1%
}{2}\right) }\text{ }_{1}F_{1}\left[ k,2k;i2r\lambda \right] ,  \notag \\
R_{2}(r) &=&BJ_{k+\frac{1}{2}}\left( \lambda r\right) =B\frac{e^{-i\lambda
r}\left( \frac{\lambda r}{2}\right) ^{k+\frac{1}{2}}}{\Gamma \left( k+\frac{3%
}{2}\right) }\text{ }_{1}F_{1}\left[ k+1,2k+2;i2r\lambda \right] .
\label{Equation18}
\end{eqnarray}%
On the other hand, to solve Eqs.(\ref{Equation16}), we must define a new
variable such as $z=\frac{\lambda }{H}e^{-Ht}$. With the definition $n=-i%
\frac{m}{H}$, the solutions of the Eqs.(\ref{Equation16}) are
obtained in terms of Bessel functions or confluent hypergeometric
functions;
\begin{eqnarray}
T_{1}\left( z\right)  &=&Cz^{\frac{3}{2}}J_{n+\frac{1}{2}}\left( z\right)
=Cz^{\frac{3}{2}}\frac{e^{-iz}\left( \frac{z}{2}\right) ^{n+\frac{1}{2}}}{%
\Gamma \left( n+\frac{3}{2}\right) }\text{ }_{1}F_{1}\left[ n+1,2n+2;i2z%
\right]   \notag \\
T_{2}\left( z\right)  &=&Dz^{\frac{3}{2}}J_{n-\frac{1}{2}}\left( z\right)
=Dz^{\frac{3}{2}}\frac{e^{-iz}\left( \frac{z}{2}\right) ^{n-\frac{1}{2}}}{%
\Gamma \left( n+\frac{1}{2}\right) }\text{ }_{1}F_{1}\left[ n,2n;i2z\right]
\label{Equation20}
\end{eqnarray}
Then, the wave function, $\Psi(x)$, can be written as
\begin{equation}
\Psi =Nz^{\frac{3}{2}}e^{ik\phi }\binom{J_{n+\frac{1}{2}}\left( z\right)
J_{k-\frac{1}{2}}(\lambda r)}{J_{n-\frac{1}{2}}\left( z\right) J_{k+\frac{1}{%
2}}(\lambda r)}  \label{Equation21}
\end{equation}%
where $N$ is normalization constant. To find the normalization constant, we
use the Dirac-delta normalization condition \cite{4,25,25a}:
\begin{equation}
\int \sqrt{\left\vert g\right\vert }\overline{\Psi }\overline{\sigma }
^{0}\Psi d\sigma =\delta \left( \lambda -\lambda ^{^{\prime }}\right)
\label{Equation22}
\end{equation}%
where $g$ is determinant of the metric tensor, $\overline{\Psi}=\Psi^{\dag }\overline{\sigma}^{0}$ and $d\sigma=drd\phi$ for the surface $t=constant$ \cite{23a}. Thus, the normalization constant is computed as
\begin{equation*}
\left\vert N\right\vert ^{2}=\frac{H^{2}}{4\pi z\lambda J_{n-\frac{1}{2}%
}\left( z\right) J_{n+\frac{1}{2}}\left( z\right) },
\end{equation*}%
where we use the following relation \cite{4,25,25a}
\begin{equation*}
\int_{0}^{\infty }rJ_{k+\frac{1}{2}}(\lambda r)J_{k+\frac{1}{2}}(\lambda
^{^{\prime }}r)dr=\frac{1}{\lambda}\delta\left(\lambda^{^{\prime}}-\lambda\right).
\end{equation*}

\section{Dirac particle in the 2+1 dimensional exponentially expanding magnetized universe}

It is interesting in discussing if the universe is
under influence in an external constant magnetic field in the beginning
time. Therefore, an electromagnetic potential can be chose as $A_{0}=0$ and $%
\overrightarrow{A}(r,\phi )=\frac{1}{2}B_{0}r\overset{\wedge }{\phi }$ for a
constant magnetic field in 2+1 dimensional spacetime. Then, the Dirac
equation in the 2+1 dimensional exponential expanding universe with a
constant magnetic field becomes
\begin{equation*}
\big\{\overline{\sigma }^{0}\left( \partial _{t}+H\right) +im+e^{-Ht}[%
\overline{\sigma }^{1}(\partial _{r}+\frac{1}{2r})+\overline{\sigma }%
^{2}\left( \frac{1}{2r}\partial _{\phi }+\frac{ieB_{0}}{2}\right) ]\big\}%
\Psi (x)=0
\end{equation*}%
and, thus, the explicit form of the equation is written as
\begin{eqnarray}
\left( \partial _{t}+im+H\right) \psi _{1}+ie^{-Ht}\left( \partial _{r}+%
\frac{1}{2r}-\frac{i}{r}\partial _{\phi }+\frac{eB_{0}}{2}\right) \psi _{2}
&=&0  \notag \\
\left( \partial _{t}-im+H\right) \psi _{2}-ie^{-Ht}\left( \partial _{r}+%
\frac{1}{2r}+\frac{i}{r}\partial _{\phi }-\frac{eB_{0}}{2}\right) \psi _{1}
&=&0  \label{Equation24}
\end{eqnarray}%
To solve the Eqs(\ref{Equation24}), we use the same procedure as section
before. The solutions of the equations are
\begin{equation}
\Psi =\frac{Nz^{\frac{3}{2}}e^{ik\phi }}{\sqrt{\rho}}\binom{J_{n+\frac{1}{2}}\left( z\right)
W_{\kappa ,\eta }(\rho )}{J_{n-\frac{1}{2}}\left(
z\right)[\frac{\left( \rho -2\eta -1\right) \left(
\sqrt{e^{2}B_{0}^{2}-4\lambda ^{2}}-eB_{0}\right) }{2\lambda \rho }W_{\kappa
,\eta }(\rho )-\frac{\sqrt{e^{2}B_{0}^{2}-4\lambda ^{2}}}{\lambda \rho }%
W_{\kappa +1,\eta }(\rho )]}  \label{Equation28}
\end{equation}%
where $\kappa =-\frac{ekB_{0}}{\sqrt{e^{2}B_{0}^{2}-4\lambda
^{2}}}$, $\eta =k-\frac{1}{2}$, $\rho =\sqrt{e^{2}B_{0}^{2}-4\lambda
^{2}}r,$ and $N$ is normalization constant. As all the contributions
for particle creation and dipole moments are taking place from the
boundaries, we can write the wave function in the following
asymptotic form \cite{25}:
\begin{equation}
\Psi \sim Nz^{\frac{3}{2}}e^{ik\phi }\rho ^{\kappa -\frac{1}{2}}e^{-\frac{%
\rho }{2}}\binom{\frac{1}{\Gamma \left( n+\frac{3}{2}\right) }\left( \frac{z%
}{2}\right) ^{n+\frac{1}{2}}}{\frac{1}{\Gamma \left( n+\frac{1}{2}\right) }%
\left( \frac{z}{2}\right) ^{n-\frac{1}{2}}\frac{\left[ \left( \rho -2\eta
-1\right) \left( \sqrt{e^{2}B_{0}^{2}-4\lambda ^{2}}-eB_{0}\right) -2\rho
\sqrt{e^{2}B_{0}^{2}-4\lambda ^{2}}-eB_{0}\right] }{2\lambda \rho }}.
\label{Equation29}
\end{equation}%
Then, the normalization constant can be obtained from the Eqs.(\ref%
{Equation22}) as follows:
\begin{equation*}
\left\vert N\right\vert ^{2}\sim \frac{H^{2}\left( \eta +1/2\right)
^{2}\left( 1-4n^{2}\right) }{\cos \left( n\pi \right) }\frac{1}{\left[
z^{2}\left( \left( \eta +1/2\right) ^{2}-\kappa ^{2}\right) \Gamma \left(
2\kappa \right) +\left( 1-4n^{2}\right) \Gamma \left( 2\kappa -2\right) %
\right] },
\end{equation*}%
where we use $\Gamma \left( -n+\frac{3}{2}\right) \Gamma \left( n+\frac{3}{2}%
\right) =\frac{\pi (1-4n^{2})}{4\cos (n\pi )}$ and $\Gamma \left( -n+\frac{1%
}{2}\right) \Gamma \left( n+\frac{1}{2}\right) =\frac{\pi }{\cos (n\pi )}$.

\section{Dirac currents}

The 2+1 dimensional Dirac current is written as
\begin{equation}
J^{\nu }=\overline{\Psi }\overline{\sigma }^{\nu }(x)\Psi  \label{Equation30}
\end{equation}%
where $\overline{\Psi }$ is hermitian conjugate of the Dirac spinor
$\Psi $ and equal to $\overline{\Psi }=\Psi ^{\dagger
}\overline{\sigma }^{0}=\Psi^{\dagger }\sigma ^{3}$ \cite{23}. As showed in \cite{23}, the Eqs.(\ref%
{Equation30}) expressed in explicit form as follow;
\begin{eqnarray}
J^{\tau } &=&\frac{1}{2m}\left( \overline{\Psi }\overline{\sigma }^{\tau
\upsilon }\left( t,r\right) \Psi \right) _{,\upsilon }-\frac{1}{2m}\overline{%
\Psi }\left( \frac{i}{2}g^{\tau \upsilon }\overleftrightarrow{\partial
_{\upsilon }}-eA^{\tau }\right) \Psi -\frac{i}{4m}\overline{\Psi }\left[
\overline{\sigma }^{\upsilon }\left( t,r\right) ,\overline{\sigma }%
_{,\upsilon }^{\tau }\left( t,r\right) \right] \Psi  \notag \\
&&-\frac{i}{2m}\overline{\Psi }\left[ \overline{\sigma }^{\upsilon }\left(
t,r\right) \Gamma _{\upsilon },\overline{\sigma }^{\tau }\left( t,r\right) %
\right] \Psi -\frac{i}{4m}\overline{\Psi }\left[ \overline{\sigma }%
_{,\upsilon }^{\upsilon }\left( t,r\right) ,\overline{\sigma }^{\tau }\left(
t,r\right) \right] \Psi  \label{Equation31}
\end{eqnarray}%
The components of the Dirac current in the 2+1dimensional exponential
expanding universe, $J^{0}$ and $J^{k}$, are
\begin{equation}
J^{0}=\frac{1}{2m}\partial _{k}\left[ \overline{\Psi }\overline{\sigma }%
^{0k}\left( t,r\right) \Psi \right] -\frac{1}{2m}\overline{\Psi }\left(
\frac{i}{2}\overleftrightarrow{\partial ^{0}}-qA^{0}\right) \Psi -\frac{i}{%
2mr}\exp \left( -Ht\right) \overline{\Psi }\overline{\sigma }^{1}\overline{%
\sigma }^{0}\Psi  \label{Equation32}
\end{equation}%
and
\begin{eqnarray}
J^{k} &=&\frac{1}{2m}\partial _{0}\left( \overline{\Psi }\overline{\sigma }%
^{0k}\left( t,r\right) \Psi \right) +\frac{1}{2m}\partial _{l}\left(
\overline{\Psi }\overline{\sigma }^{lk}\left( t,r\right) \Psi \right) -\frac{%
1}{2m}\overline{\Psi }\left( \frac{i}{2}\overleftrightarrow{\partial ^{k}}%
-qA^{k}\right) \Psi  \notag \\
&&+i\frac{3H}{2mr}\exp \left( -Ht\right) \overline{\Psi }\overline{\sigma }%
^{0}\overline{\sigma }^{k}\Psi +\delta _{k2}i\frac{3H}{2mr}\exp \left(
-Ht\right) \overline{\Psi }\overline{\sigma }^{1}\overline{\sigma }^{2}\Psi
\label{Equation33}
\end{eqnarray}%
where $k,l=1,2$, $\overline{\sigma}^{0k}=i/2\left[ \overline{\sigma }^{0},%
\overline{\sigma }^{k}\left( t,r\right) \right] $, $\overline{\sigma }%
^{kl}=i/2\left[ \overline{\sigma }^{k}\left( t,r\right) ,\overline{\sigma }%
^{l}\left( t,r\right) \right] $ and $\delta _{k2}$ is Dirac-delta function.
Also this components can be rewritten in terms of the convective, the
polarization and magnetization parts as follows:%
\begin{equation*}
J_{0}=\partial _{k}\mathbf{P}_{k}+\rho _{convective}-\frac{i}{2mr}\exp
\left(-Ht\right) \overline{\Psi }\overline{\sigma }^{1}\overline{\sigma }%
^{0}\Psi
\end{equation*}%
and
\begin{equation*}
J_{k}=\partial _{0}\mathbf{P}_{k}+\partial _{l}\mathbf{M}_{\left[ lk\right]
}+J_{k~convective}+i\frac{3H}{2mr}\exp \left( -Ht\right) \overline{\Psi }%
\overline{\sigma }^{0}\overline{\sigma }^{k}\Psi +\delta _{k2}i\frac{3H}{2mr}%
\exp \left( -Ht\right) \overline{\Psi }\overline{\sigma }^{1}\overline{%
\sigma }^{2}\Psi
\end{equation*}%
where $\mathbf{P}_{0k}$ are polarization densities and $\mathbf{M}_{\left[ lk%
\right] }$ is magnetization density, and their explicit forms are given by,
\begin{equation}
P^{0k}=\frac{1}{2m}\overline{\Psi }\overline{\sigma }^{k0}(t,r)\Psi
\label{Equation34}
\end{equation}%
and
\begin{equation}
M^{[lk]}=\frac{1}{2m}\overline{\Psi }\overline{\sigma }^{lk}(t,r)\Psi ,
\label{Equation35}
\end{equation}%
respectively. From these relations, the total polarizations, $p_{l}^{0},$
and magnetization, $\mu ,$ are defined as
\begin{equation}
p_{l}^{0}=\int P^{0k}d\Sigma _{kl}  \label{Equation36}
\end{equation}%
and
\begin{equation}
\mu =\int M^{kl}d\Sigma _{kl},  \label{Equation37}
\end{equation}%
where $d\Sigma _{kl}$ is an hypersurface for $t=$constant and $d\Sigma _{kl}=%
\sqrt{\left\vert g\right\vert }d^{2}x=e^{2Ht}rdrd\phi$ \cite{23a}.

Now, we are going to discuss the Dirac currents and the dipole moments
expressions for the exponentially expending universe. So, inserting the Eqs.(%
\ref{Equation21}) and its conjugate into the Eqs.(\ref{Equation30}), we
compute the components of the Dirac currents in asymptotic region as follow:
\begin{eqnarray}
J^{0} &\approx &\frac{H^{2}}{2\lambda ^{2}\pi ^{2}}\frac{z^{2}}{r},  \notag
\\
J^{1} &\approx &\frac{H^{3}z^{2}\left[ \left( 2n+1\right) ^{2}-z^{2}\right]
}{4\pi ^{2}\lambda ^{3}\left( 2n+1\right) r}\sin \left( \lambda r-\frac{k\pi
}{2}\right) ,  \notag \\
J^{2} &\approx &i\frac{H^{3}z^{2}\left[ z^{2}-\left( 2n+1\right) ^{2}\right]
}{4\pi ^{2}\lambda ^{3}\left( 2n+1\right) r}\ \sin \left( \lambda r-\frac{%
k\pi }{2}\right) .  \label{Equation38}
\end{eqnarray}%
Similarly, substituting Eqs.(\ref{Equation21}) and its conjugate in the Eqs.(%
\ref{Equation34}) and Eqs.(\ref{Equation35}), the components of the
polarization densities and the magnetization density are written as
follows:
\begin{eqnarray}
P^{1} &=&A\left( z\right) J_{k+\frac{1}{2}}(\lambda r)J_{k-\frac{1}{2}%
}(\lambda r),  \notag \\
P^{2} &=&B\left( z\right) \frac{1}{r}J_{k+\frac{1}{2}}(\lambda r)J_{k-\frac{1%
}{2}}(\lambda r)\ ,  \notag \\
M_{12} &=&\frac{\lambda ^{4}r}{8\pi m^{3}}\left[ J_{k+\frac{1}{2}}(\lambda
r)J_{k+\frac{1}{2}}(\lambda r)-J_{k-\frac{1}{2}}(\lambda r)J_{k-\frac{1}{2}%
}(\lambda r)\right] ,  \label{Equation39}
\end{eqnarray}%
respectively, where we use the following abbreviations:
\begin{eqnarray*}
A\left( z\right)  &=&i\frac{H^{3}z^{3}}{8\pi \lambda ^{2}m}\left[ \frac{J_{n+%
\frac{1}{2}}(z)}{J_{n-\frac{1}{2}}(z)}-\frac{J_{n-\frac{1}{2}}(z)}{J_{n+%
\frac{1}{2}}(z)}\right] , \\
B\left( z\right)  &=&-iA\left( z\right) .
\end{eqnarray*}%
Giving to the polarization densities and magnetization density depending
spacetime coordinates, we can say that particle production event takes
place. To calculate the total polarizations and magnetization, we insert the
Eq.(\ref{Equation39}), the polarization densities and magnetization
density, in the Eqs.(\ref{Equation36}) and (\ref{Equation37}),
respectively, and later integrate them on the hypersurface. Then, we obtain
the total polarization densities (electric dipole moments) and magnetization
density (magnetic dipole moment) as follows:
\begin{eqnarray*}
p^{1}&=&0,
\notag \\
p^{2}&=&\frac{\pi \lambda }{H^{2}}\frac{B\left( z\right) }{z^{2}},
\notag \\
\mu &=&0,
\end{eqnarray*}%
where we use the integral representation of Bessel function \cite{25} and $%
p^{1}$ and $p^{2}$ are total polarizations, i.e. electric dipole moment
components, and also $\mu $ is total magnetization, i.e. magnetic dipole
moment. From these results, we see that the particle creation events are
affected from only $p^{2}$ total polarization density, electric dipole
moment. On the other hand, in the limit $t\rightarrow -\infty \left(
z\rightarrow \infty \right) $, $p^{2}$ vanishes by the following way
\begin{equation*}
p^{2}\rightarrow \frac{zH}{4\lambda m}\left[ \frac{2\cos ^{2}\left( z-\frac{%
n\pi }{2}\right) -1}{\sin \left( 2z-n\pi \right) }\right] \rightarrow 0,
\end{equation*}%
that is, there is not particle production in this limit or the universe has
a symmetric charge distribution in the beginning time and, of course, the
universe has not any dipole moments, but, in the limit $t\rightarrow +\infty
\left( z\rightarrow 0\right) $, $p^{2}$ becomes
\begin{equation}
p^{2}\rightarrow \frac{e\hbar H}{4\lambda mc^{2}}\left[ \frac{z^{2}-\left(
1-i\frac{2mc^{2}}{\hbar H}\right) ^{2}}{1-i\frac{2mc^{2}}{\hbar H}}\right]
\rightarrow -\frac{e}{2\lambda\delta}\exp\left(-i\delta\right),
\label{Equation39a}
\end{equation}%
where $\hbar $ is\ Planck constant, $c$ \ is the speed of light and $\delta =%
\frac{2mc^{2}}{\hbar H},$ and, thus, the universe has a permanent complex
dipole moment, which it oscillates with Zitterbewegung frequency, $\frac{%
2mc^{2}}{\hbar }.$

To calculate the Dirac current components for the Dirac particle in the 2+1
dimensional exponentially expanding magnetized universe, we insert the Eqs.(%
\ref{Equation29}) and its conjugate in the Eqs.(\ref{Equation30}). Thus, we
find that the current components are%
\begin{eqnarray}
J^{0} &\approx &\frac{H^{2}k^{2}}{z^{2}\left( k^{2}-\kappa ^{2}\right)
\Gamma \left( 2\kappa \right) +e^{2}B_{0}^{2}\left( 1-4n^{2}\right) \Gamma
\left( 2\kappa -2\right) }z^{3}\rho ^{2\kappa -1}e^{-\rho }  \notag \\
&&\left[ z+\frac{\left( 1-4n^{2}\right) }{z\left( k^{2}-\kappa ^{2}\right)
\rho ^{2}}\left[ \left( 2k-\rho \right) \left( k+\kappa \right) +2k\rho %
\right] ^{2}\right]   \notag \\
J^{1} &\approx &i\frac{4nH^{3}eB_{0}k^{2}\kappa }{\pi \left( k^{2}-\kappa
^{2}\right) }z^{4}\rho ^{2\kappa -2}e^{-\rho }\left[ \left( 2k-\rho \right)
\left( k+\kappa \right) +2k\rho \right]   \notag \\
J^{2} &\approx &\frac{4H^{3}k^{3}z^{4}}{\pi \left( k^{2}-\kappa ^{2}\right) }%
\frac{\left( \rho -2k\right) \left( k+\kappa \right) +2k\rho }{z^{2}\left(
k^{2}-\kappa ^{2}\right) \Gamma \left( 2\kappa \right) +\left(
1-4n^{2}\right) \Gamma \left( 2\kappa -2\right) }\rho ^{2\kappa -3}e^{-\rho
}.  \label{Equation40}
\end{eqnarray}%
Using Eqs.(\ref{Equation29}) in the Eqs.(\ref{Equation34}) and in the Eqs.(%
\ref{Equation35}), the components of polarization and magnetization can be
found as follows:
\begin{eqnarray}
P^{1} &\approx &\frac{2H^{3}eB_{0}k^{2}z^{4}\kappa }{\left( k^{2}-\kappa
^{2}\right) \pi m}\frac{\left( 2k-\rho \right) \left( k+\kappa \right)
+2k\rho }{2z^{2}\lambda ^{2}\kappa ^{2}\Gamma \left( 2\kappa \right)
+e^{2}B_{0}^{2}\left( 1-4n^{2}\right) \Gamma \left( 2\kappa -2\right) }\rho
^{2\kappa -2}e^{-\rho }  \notag \\
P^{2} &\approx &i\frac{4nH^{3}k^{3}z^{4}}{\pi m\left( k^{2}-\kappa
^{2}\right) }\frac{\left[ \left( 2k-\rho \right) \left( k+\kappa \right)
+2k\rho \right] }{z^{2}\left( k^{2}-\kappa ^{2}\right) \Gamma \left( 2\kappa
\right) +\left( 1-4n^{2}\right) \Gamma \left( 2\kappa -2\right) }\rho
^{2\kappa -3}e^{-\rho }  \notag \\
M_{12} &\approx &\frac{2H^{4}eB_{0}k^{3}z^{5}\left( 4n^{2}-1\right) \kappa }{%
\pi m\left( k^{2}-\kappa ^{2}\right) \left[ 4z^{2}\lambda ^{2}\kappa
^{2}\Gamma \left( 2\kappa \right) +eB_{0}\left( 1-4n^{2}\right) \Gamma
\left( 2\kappa -2\right) \right] }\rho ^{2\kappa -2}e^{-\rho }  \notag \\
&&\times\left[ \frac{z}{\left( 1-4n^{2}\right) }+\frac{1}{z\left( k^{2}-\kappa
^{2}\right) \rho ^{2}}\left[ \left( 2k-\rho \right) \left( k+\kappa \right)
+2k\rho \right] ^{2}\right]   \label{Equation41}
\end{eqnarray}%
To calculate the the total polarizations and total magnetization, we insert
the Eqs.(\ref{Equation41}) in Eqs.(\ref{Equation36}) and
Eqs.(\ref{Equation37}). Then, using the integral representation of
Bessel function \cite{25}, we find the following dipole moments
expressions:
\begin{eqnarray*}
p^{1}&\approx & \frac{He^{3}B_{0}^{3}k^{4}z^{2}}{\kappa ^{3}m}\frac{2\left(
k-\kappa \right) ^{2}\Gamma \left( 2\kappa \right) }{z^{2}\left(
k^{2}-\kappa ^{2}\right) \Gamma \left( 2\kappa \right) +\left(
1-4n^{2}\right) \Gamma \left( 2\kappa -2\right) },
\notag \\
p^{2} &\approx& i\frac{2nHe^{4}B_{0}^{4}k^{5}z^{2}}{\kappa ^{4}m}\frac{\left(
2k^{2}-2\kappa ^{2}+4k\kappa +\kappa -k\right) \Gamma \left( 2\kappa
-1\right) }{z^{2}\left( k^{2}-\kappa ^{2}\right) \Gamma \left( 2\kappa
\right) +\left( 1-4n^{2}\right) \Gamma \left( 2\kappa -2\right) },
\notag \\
\mu &\approx& \frac{H^{2}e^{3}B_{0}^{3}k^{5}}{\left( k^{2}-\kappa ^{2}\right)
\kappa ^{3}m}z^{3}.
\end{eqnarray*}
From these expressions, we see that, in finite time intervals, the particle
creation influenced by both polarization and magnetization components. On
the other hand, in the limit $t\rightarrow \infty \left( z\rightarrow
0\right) $ the magnetic dipole moment, $\mu ,$ goes to zero faster than the
electric dipole moment components, $p^{1}$ and $p^{2}$. Therefore, the
electric dipole moments in the particle creation events become more dominant
than the magnetic dipole moment in finite time intervals if there exists in
an external constant magnetic field.

\section{Summary and Conclusion}

We exactly solve the Dirac equation in existence of the exponentially
expanding magnetized and non-magnetized universe and, from these solutions,
derive some expressions for the Dirac current components and dipole moments.
The particle creation in the exponentially expanding universe are only
affected by the $p^{2}$ polarization in the finite time interval. However,
this component goes to zero in the limit $t\rightarrow -\infty $, i.e. in
the beginning of the universe, but, in the limit $t\rightarrow +\infty ,$
the universe has a permanent complex dipole moment oscillating with
Zitterbewegung frequency, $\frac{2mc^{2}}{\hbar }$: $p^{2}\simeq $ $-\frac{e%
}{2\lambda \delta }\exp \left(-i\delta \right)$. The complexity of the
dipole moment points out the conductivity of the exponentially expanding
universe. Also, the universe has the electric and magnetic dipole moments
which are dependent on time in existence of an external constant magnetic
field with the expansion such that, in the limit $t\rightarrow-\infty $,
the dipole moment expressions become infinite, but, in the limit $%
t\rightarrow \infty$, they go to zero. The dependence on time of the
polarization and magnetization show that the particle creation happens.
Furthermore, in the limit $t\rightarrow \infty $ $\left( z\rightarrow
0\right) $, the particle creation events are affected only\ via the
polarization because the magnetization, $m,$ goes to zero faster than $p^{1}$
and $p^{2}$. From the point of view, we point out that the exponential
expansion of the universe causes a particle creation, a permanent complex
electric dipole moment and asymmetric charge distribution$,$ but, in
existence of an external constant magnetic field with exponential expansion
in time, the universe charge distribution is get and getting symmetric and
thus all the dipole moments become zero as $t\rightarrow \infty $.

\section*{Conflict of Interests}

The authors declare that there is no conflict of interests regarding the publication of this paper.

\section*{Data Availability}

No data were used to support this study.
\section*{Funding Statement}
No funding were used to support this study.

%\newpage
%\bibliography{apssamp}% Produces the bibliography via BibTeX.

\clearpage

\end{document}